\UseRawInputEncoding
\documentclass[twocolumn,showpacs,preprintnumbers,superscriptaddress,amsmath,amssymb]{revtex4-1}
\usepackage{wallpaper}
\usepackage{graphicx}
\usepackage{dcolumn}
\usepackage{bm}
\usepackage{color}
\usepackage{multirow}
\usepackage{CJK}
\usepackage{mathptmx}
\usepackage[colorlinks,linkcolor=blue,anchorcolor=blue,citecolor=blue]{hyperref}

\begin{document}
		\renewcommand\arraystretch{1.2}
		
		\title{$0\nu\beta\beta$-decay nuclear matrix elements in self-consistent Skyrme quasiparticle random phase approximation: uncertainty from pairing interaction}
		
		\author{W.-L. Lv}
		\address{School of Nuclear Science and Technology, Lanzhou University, Lanzhou 730000, China}
		\address{Frontiers Science Center for Rare isotopes, Lanzhou University, Lanzhou 730000, China}
		\author{Y.-F. Niu}\email{niuyf@lzu.edu.cn}
		\address{School of Nuclear Science and Technology, Lanzhou University, Lanzhou 730000, China}
		\address{Frontiers Science Center for Rare isotopes, Lanzhou University, Lanzhou 730000, China}
		\author{D.-L. Fang}
		\address{Institute of Modern Physics, Chinese Academy of Sciences, Lanzhou 730000, China}
		\author{J.-M. Yao}
		\address{School of Physics and Astronomy, Sun Yat-sen University, Zhuhai, 519082, China}
		\author{C.-L. Bai}
		\address{Department of Physics, Science, and Technology, Sichuan University, Chengdu 610065, China}
        \author{J. Meng}
		\address{School of Physics, Peking University, Beijing 100871, China}

		
		
%
		\begin{abstract}
			The uncertainty in the nuclear matrix elements (NMEs) of $0\nu\beta\beta$ decay
			for $^{76}$Ge, $^{82}$Se, $^{128}$Te, $^{130}$Te, and $^{136}$Xe in
			the self-consistent quasiparticle random phase approximation
			(QRPA) method is investigated by using eighteen Skyrme interactions supplemented
			with either a volume- or surface-type of pairing interactions.
			The NMEs for the isotopes concerned (except $^{136}$Xe)
            are less sensitive to the particle-hole ($ph$) interactions,
			while strongly dependent on the employed isovector particle-particle ($pp$) pairing interactions
			even though the pairing strengths are optimized to the same pairing gap.
			The results indicate that a precise determination of the isovector $pp$ pairing interaction
			in the Skyrme energy density functional is of importance to reduce the uncertainty in the NMEs within the QRPA framework.
		\end{abstract}

		\maketitle
		
		
	\section{Introduction}
	\label{secIntro}
	
	Neutrinoless double-beta ($0\nu\beta\beta$) decay is a lepton-number-violating process,
	which is preferred by extensions of the standard model \cite{Avignone2008,Ejiri2019,Dolinski2019}.
	Such a hypothetical second-order transition occurs only if neutrinos are their own antiparticles, i.e., the Majorana particles.
	Besides demonstrating the existence of lepton-number-violation,
	if this process is driven by the standard mechanism of exchange light Majorana neutrinos, the discovery of $0\nu\beta\beta$ decay
	provides a practical way to determine the mass scale and mass hierarchy of neutrinos,
	provided that the corresponding nuclear matrix elements (NMEs) $M^{0\nu}$ are known.
	Since not only the unknown neutrino properties but also nuclear physics are involved,
	the NMEs can only be obtained by nuclear many-body approaches,
	see for instance the recent reviews~\cite{Vergados2012,Engel2017,Yao2022},
	among which quasiparticle random phase approximation (QRPA) and interacting shell model (ISM) are two widely used microscopic models.
	
	Due to the large discrepancy in the NMEs obtained by different nuclear models,
	a great deal of efforts has been made to reduce this uncertainty.
	Based on the nuclear models of either ISM \cite{Menendez2009} or
	QRPA with $G$-matrix-based residual interactions ($G$-QRPA) \cite{Simkovic1999,Rodin2003,Rodin2006a,Simkovic2008,Kortelainen2007,Suhonen2008},
	the effects of the induced currents, the size of the single-particle basis, dipole-form-factor cut-off parameters, and short-range correlation
	have been assessed.
	In particular, by using different realistic nucleon-nucleon ($NN$) potentials renormalized by $G$ matrix, including Bonn, Argonne, and Nijmegen,
	Rodin {\it et al.} found the NMEs are essentially independent of the residual interactions in the QRPA \cite{Rodin2003}.
	However, in their QRPA approach the ground state is obtained by the same Coulomb corrected Woods-Saxon potential,
	even though the residual interactions for excited states are different.
	Therefore, the uncertainty study of $M^{0\nu}$ using $NN$ interactions needs to be revisited with the self-consistent QRPA,
	where the same effective $NN$ interactions are consistently used for both the ground state and excited states.
	
	In the past decade, the fully self-consistent QRPA based on Skyrme energy density functionals (EDFs) has been applied to the study of $\beta\beta$ decay.
	The effects of overlap factor of excited states, many-body correlations, as well as the isoscalar pairing on $M^{0\nu}$
	are studied \cite{Terasaki2012,Terasaki2015,Terasaki2019,Terasaki2020}.
	In Ref.~\cite{Mustonen2013}, axially deformed Skyrme QRPA was used to calculate $M^{0\nu}$ for the first time.
	Recently, the NMEs of $2\nu\beta\beta$ decay, $M^{2\nu}$, calculated by  deformed Skyrme QRPA using finite amplitude method \cite{Hinohara2022},
	and spherical relativistic QRPA \cite{Popara2022} became available.
	In these studies, however, the uncertainty of $M^{0\nu}$ was not investigated.
	
	In this work, we focus on the uncertainty induced by the particle-hole ($ph$) channel and particle-particle ($pp$) channel
	of effective interactions with a fully self-consistent QRPA based on Skyrme EDFs.
	A brief introduction of the formalism is outlined in Sec.~\ref{secTheo}.
	In Sec.~\ref{secResu}, we present the results of $M^{0\nu}$ calculated by different $ph$ and $pp$ interactions.
	Conclusions and perspectives are given in Sec.~\ref{secConclu}.
	
	\section{Formalism}
	\label{secTheo}
	
	In the self-consistent QRPA, the same effective nuclear interaction is employed to solve the Hartree-Fock-Bogoliubov (HFB) equation
	and the follow-up QRPA equation.
	For the $ph$ channel, we use the Skyrme interaction. For the $pp$ channel, we use the $\delta$ interaction \cite{Dobaczewski2002,Bennaceur2005},
	\begin{equation}
		V^{pp}(\bm{r}_1, \bm{r}_2) = \left[ t'_0 + \frac{t'_3}{6}\rho(\frac{\bm{r}_1+\bm{r}_2}{2}) \right] \delta(\bm{r}_1 - \bm{r}_2),
	\end{equation}
	which is referred to the surface- and volume-type of pairing interactions for the $t'_3$ term is switched on or off.
	In the case of surface pairing,  $t'_3=-37.5 t'_0$ is employed,
	where the pairing field is peaked at the nuclear surface and follows roughly the variations of the nucleon density.
	
	The nuclear matrix element of $0\nu\beta\beta$ decay is defined as
	\begin{equation}
		M^{0\nu} \equiv -M^{0\nu}_{{\rm F}} + M^{0\nu}_{{\rm GT}} + M^{0\nu}_{{\rm T}},
	\end{equation}
	where the tensor term $M^{0\nu}_{{\rm T}}$ is negligibly small \cite{Simkovic1999,Mustonen2013}.
	For the ground-state-to-ground-state transition $(0^{(i)+}_{\rm g.s.} \rightarrow 0^{(f)+}_{\rm g.s.})$, the Fermi term is
	\begin{equation}
		\begin{aligned}
			M^{0\nu}_{\rm F}
			&= \displaystyle \frac{8R}{g_A^{2}}
			\int q^2 \mathrm{d}q \sum_{N_i N_f} \sum_{J^P}
			\sum_{\pi_i \nu_i} \sum_{\pi_f \nu_f}  \hat{J}^{-2}
			\frac{h_{\rm F}(q^2) \langle N_f J^P|N_i J^P \rangle }{  q(q + E_d ) } \\
			&  \times
			\langle 0^{(f)+}_{\rm g.s.}|| [c^{\dag}_{\pi_f}  \tilde{c}_{\nu_f}]_{J} ||N_f J^P\rangle
			\langle N_i J^P|| [c^{\dag}_{\pi_i}  \tilde{c}_{\nu_i}]_{J} ||0^{(i)+}_{\rm g.s.}\rangle \\
			&  \times (-)^{J}
			\langle j_{\pi_f} || j_{J}(qr) Y_{J}|| j_{\nu_f} \rangle
			\langle j_{\pi_i} || j_{J}(qr) Y_{J}|| j_{\nu_i} \rangle , \\
		\end{aligned}
	\end{equation}
	and the GT term is
	\begin{equation}
		\begin{aligned}
			& M^{0\nu}_{\rm GT}
			= \displaystyle \frac{8 R}{g_A^2}
			\int q^2 \mathrm{d}q \sum_{N_i N_f} \sum_{J^P} \sum_{\pi_i \nu_i} \sum_{\pi_f \nu_f} \hat{J}^{-2}
			\frac{ h_{\rm GT}(q^2) \langle N_f J^P|N_i J^P \rangle }{q(q + E_d)}  \\
			& \times
			\langle 0^{(f)+}_{\rm g.s.}|| [c^{\dag}_{\pi_f}  \tilde{c}_{\nu_f}]_{J} ||N_f J^P\rangle
			\langle N_i J^P|| [c^{\dag}_{\pi_i}  \tilde{c}_{\nu_i}]_{J} ||0^{(i)+}_{\rm g.s.}\rangle \\
			& \times  \! \sum_{l=J-1}^{J+1} (-)^{l}
			\langle j_{\pi_f} || j_{l}(qr) [Y_{l} \sigma]_{J} || j_{\nu_f} \rangle
			\langle j_{\pi_i} || j_{l}(qr) [Y_{l} \sigma]_{J} || j_{\nu_i} \rangle .
		\end{aligned}
	\end{equation}
	Here $j_{k}(qr)$ is the spherical Bessel function of $k$-th order,
	and $Y_k$ is the spherical harmonic function. An empirical formula of the nuclear radius is adopted, i.e., $R=1.2A^{1/3}$fm.
	In QRPA model, the closure approximation can be avoided,
	$E_d = (\Omega_{N_i} + \Omega_{N_f} + \Delta\lambda_{\pi \nu}^{(i)}
	- \Delta\lambda_{\pi \nu}^{(f)})/2$,
	where $\Omega_{N_i}$ ($\Omega_{N_f}$) and $\Delta\lambda_{\pi \nu}^{(i)}$ ($\Delta\lambda_{\pi \nu}^{(f)}$)
	are respectively the eigenvalues of QRPA equation and the difference of proton and neutron Fermi surfaces
	for mother (daughter) nucleus [cf. Eq(10) in Ref.~\cite{Lv2022}].
	$\langle N_f J^P|N_i J^P \rangle$ is the overlap factor of the two QRPA excited states \cite{Rodin2006}.
	The one-body transition densities in mother and daughter nuclei are
	\begin{equation}
		\begin{aligned}
			\langle N_i J^P|| [c^{\dag}_{\pi_i}  \tilde{c}_{\nu_i}]_{J} ||0^{(i)+}_{\rm g.s.}\rangle
			=& -\hat{J}
			[ X_{\pi_i \nu_i}^{N_i J^P \ast} u_{\pi_i} v_{\nu_i}
			+        Y_{\pi_i \nu_i}^{N_i J^P \ast} v_{\pi_i} u_{\nu_i}] ,\\
			\langle 0^{(f)+}_{\rm g.s.}|| [c^{\dag}_{\pi_f}  \tilde{c}_{\nu_f}]_{J} ||N_f J^P\rangle
			=& -\hat{J}
			[ X_{\pi_f \nu_f}^{N_f J^P} v_{\pi_f} u_{\nu_f}
			+      Y_{\pi_f \nu_f}^{N_f J^P} u_{\pi_f} v_{\nu_f}  ]  ,
		\end{aligned}
		\label{Eq_1bTD}
	\end{equation}
	where $\hat{J} = \sqrt{2J+1}$. $X_{\pi \nu}^{N J^P}$ and $Y_{\pi \nu}^{N J^P}$ are respectively the forward and backward amplitudes of QRPA.
	The form factors $h_{\rm F}$ and $h_{\rm GT}$ are
	\begin{equation}
		\begin{aligned}
			h_{\rm F}(q^2) =& -g_{V}^2(q^2),  \\
			h_{\rm GT}(q^2) =& g_A^2(q^2)
			-   \frac{q^2}{3m} g_P(q^2) g_A(q^2) \\ &
			+   \frac{q^2}{6 m^2} {g}_M^{\prime 2}(q^2)
			+   \frac{q^4}{12 m^2} g_P^2(q^2),
		\end{aligned}
	\end{equation}
	with
	\begin{equation}
		\begin{aligned}
			g_V(q^2) =&~ \frac{1}{[ 1 + q^2/\Lambda_V^2 ]^2}     ,\\
			g_A(q^2) =&~ \frac{1.27}{[ 1 + q^2/\Lambda_A^2 ]^2}  ,\\
			g_P(q^2) =&~ \frac{2m g_A(q^2)}{q^2 + m_{\pi}^2}    ,\\
			g'_M(q^2)=&~ g_M(q^2) + g_V(q^2) = 4.70 g_V(q^2)  .
		\end{aligned}
	\end{equation}
	Here $m$ and $m_{\pi}$ denote the mass of proton and $\pi$ meson, respectively.
	The energy cut-offs $\Lambda^2_V=0.71$(GeV)$^2$ and $\Lambda_A^2=1.09$(GeV)$^2$ \cite{Simkovic1999}.
	
	\section{Results and discussions}
	\label{secResu}
	In Skyrme HFB calculations, the pairing (neutron-neutron and proton-proton) interactions
	are fixed to reproduce the experimental pairing gaps obtained from three-point formula of binding energies.
	For the residual pairing interactions in QRPA,
	the strengths of isoscalar proton-neutron pairing $f_{\rm IS}$ are fixed by tuning $M^{2\nu}_{\rm GT}$ to the experimental data \cite{Barabash2015},
	while the strengths of isovector channel $f_{\rm IV}$ are determined by $M^{2\nu}_{\rm F}=0$ due to the isospin symmetry \cite{Rodin2011}.
	Intermediate states of $J^P=0^{\pm} , 1^{\pm} , ..., 10^{\pm}$ are considered.
	An unquenched value of axial-vector coupling constant $g_A = 1.27$ is used.
	
	\begin{figure}[t]
		\centering
		\includegraphics[width=0.4\textwidth]{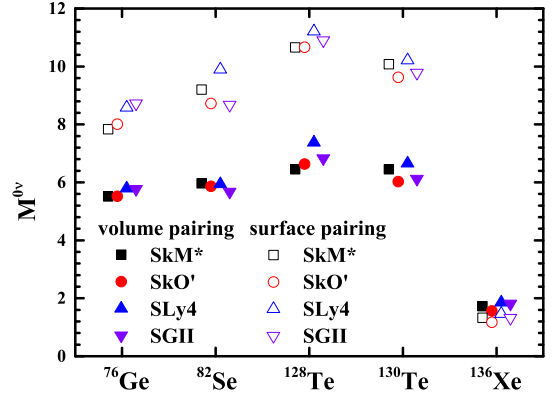}\\
		\caption{$M^{0\nu}$ for $^{76}$Ge, $^{82}$Se, $^{128}$Te, $^{130}$Te, and $^{136}$Xe, calculated by self-consistent QRPA model
			with 4 Skyrme interactions (denoted by different shapes of symbols) and two kinds of pairing interactions (denoted by solid symbols for volume pairing and hollow symbols for surface pairing).
		}\label{Fig1}
	\end{figure}
	
	We first perform a systematic calculation on the NMEs of $0\nu\beta\beta$
	for $^{76}$Ge, $^{82}$Se, $^{128}$Te, $^{130}$Te, and $^{136}$Xe,
	where SkM$^{\ast}$ \cite{Bartel1982SkMst}, SkO$'$ \cite{Reinhard1999SkO}, SLy4 \cite{Chabanat1998SLy45}, and SGII \cite{VanGiai1981SGII}
	interactions are used for the $ph$ channel.
	The tensor force is not included in this work.
	Both volume and surface pairing interactions are used for the $pp$ channel.
	Results are depicted in Fig.~\ref{Fig1}.
	One can see clearly that with the same kind of pairing interaction, $M^{0\nu}$ obtained by different $ph$ interactions are
	close to each other, with the discrepancy less than 15\% for all the nuclei of concerned.
	However, $M^{0\nu}$ is sensitive to the use of pairing interactions. The use of the surface pairing interaction leads to a much larger $M^{0\nu}$ than the use of volume pairing, except for $^{136}$Xe.
	We note that $B(E2;0^{+}_{\rm g.s.}\rightarrow 2^{+}_1)$ values
	for both mother and daughter nuclei can be better reproduced
	by volume pairing compared to surface pairing, especially for SkM$^{\ast}$ and SGII.
	
	\begin{figure}[t]
		\centering
		\includegraphics[width=0.4\textwidth]{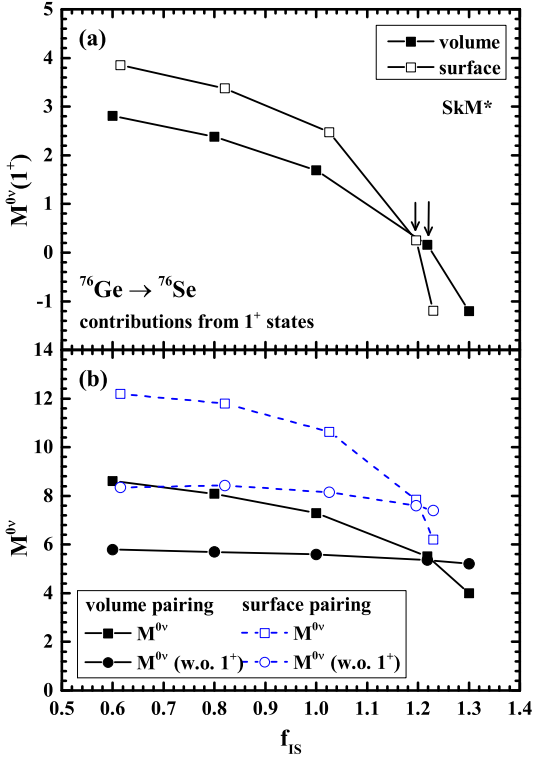}\\
		\caption{$M^{0\nu}$ for $^{76}$Ge as a function of the isoscalar pairing strength $f_{\rm IS}$.
			The contributions to $M^{0\nu}$ from $1^{+}$ intermediate states are shown in panel (a),
			where the arrows mark the $f_{\rm IS}$ fixed by experimental $M^{2\nu}_{\rm GT}$.
			The total values and the values without the contributions from $1^{+}$ intermediate states
			of $M^{0\nu}$ are shown in panel (b).}\label{Fig2}
	\end{figure}
	
	To understand the difference of $M^{0\nu}$ between volume pairing and surface pairing,
	we plot in Fig.~\ref{Fig2} the $M^{0\nu}$ for $^{76}$Ge as a function of the  isoscalar pairing  strength $f_{\rm IS}$.
	The contribution from $1^+$ states to $M^{0\nu}$, denoted as $M^{0\nu}(1^{+})$  shown in Fig. \ref{Fig2}(a), decreases rapidly with increasing $f_{\rm IS}$,
	which is similar to the trend of NMEs of $2\nu\beta\beta$ decay $M^{2\nu}_{\rm GT}$ \cite{Lv2022}.
	It comes from the fact that for both $M^{0\nu}(1^{+})$ and $M^{2\nu}_{\rm GT}$, only $1^+$ intermediate states are involved,
	which are sensitive to isoscalar pairing.
	By adjusting the value of $f_{\rm IS}$ to reproduce the experimental $M^{2\nu}_{\rm GT}$
	in the calculations, the values of $M^{0\nu}(1^{+})$ by the volume
	and surface pairing interactions are close to each other, as seen from Fig.~\ref{Fig2}(a).
	However, unlike $1^{+}$ states, other multipoles are not so sensitive to isoscalar pairing
	so that their contributions to $M^{0\nu}$ are stable for different $f_{\rm IS}$, as shown in panel (b).
	As a result, the behavior of the total $M^{0\nu}$ mainly depends on the
	the decreasing trend of $M^{0\nu}(1^{+})$ when increasing $f_{\rm IS}$,
	while the difference of $M^{0\nu}$ between volume pairing and surface pairing comes from
	the different contributions of other multipoles in these two cases.
	Since $M^{0\nu}$ contributed by other multipoles is not sensitive to isoscalar pairing, the difference in the $M^{0\nu}$
	from the different form of pairing interaction should be caused by the isovector pairing part.
	
	\begin{figure}[!t]
		\centering
		\includegraphics[width=0.4\textwidth]{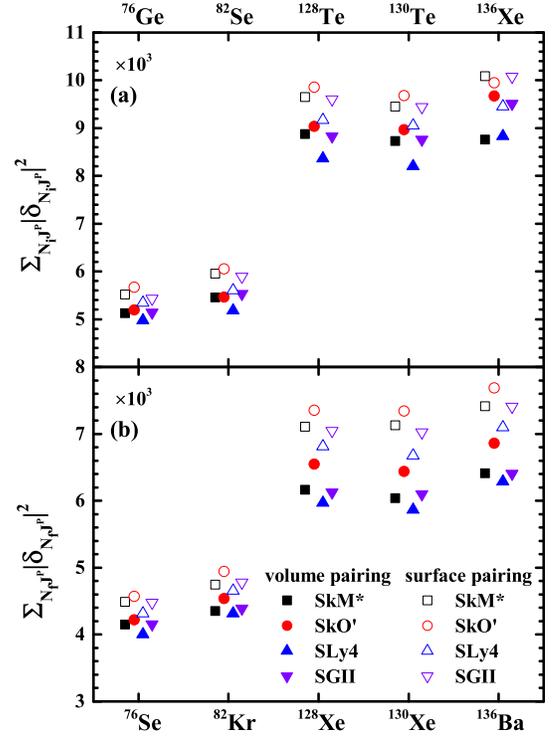}\\
		\caption{$\sum_{N J^P} |\delta_{N J^P}|^2$ in mother nuclei (a)
			and in daughter nuclei (b).}\label{Fig3}
	\end{figure}
	
	In the calculation of $M^{0\nu}$, pairing interaction plays its role mainly through the overlap of HFB wave functions
	$\langle {\rm HFB}_{f}|{\rm HFB}_{i} \rangle$,
	one-body transition densities, as well as the number of two quasiparticle (2qp) proton-neutron configurations.
	To qualitatively investigate the pairing effects on $M^{0\nu}$,
	we introduce the sum of one-body transition densities $\delta_{N J}$, defined as
	\begin{equation}
		\begin{aligned}
			\delta_{N_i J^P} \equiv& -\hat{J}^{-1} \sum_{\pi_i \nu_i} \langle N_i J^P|| [c^{\dag}_{\pi_i}  \tilde{c}_{\nu_i}]_{J} ||0^{(i)+}_{\rm g.s.}\rangle  \\
			=&  \sum_{\pi_i \nu_i}
			\left( X_{\pi_i \nu_i}^{N_i J^P \ast} u_{\pi_i} v_{\nu_i}
			+      Y_{\pi_i \nu_i}^{N_i J^P \ast} v_{\pi_i} u_{\nu_i} \right) ,\\
			\delta_{N_f J^P} \equiv& -\hat{J}^{-1} \sum_{\pi_f \nu_f} \langle 0^{(f)+}_{\rm g.s.}|| [c^{\dag}_{\pi_f}  \tilde{c}_{\nu_f}]_{J} ||N_f J^P\rangle \\
			=& \sum_{\pi_f \nu_f}
			\left( X_{\pi_f \nu_f}^{N_f J^P} v_{\pi_f} u_{\nu_f}
			+      Y_{\pi_f \nu_f}^{N_f J^P} u_{\pi_f} v_{\nu_f}  \right)  ,
		\end{aligned}
	\end{equation}
	where the occupation amplitudes are involved in one-body transition densities
	and the number of 2qp configurations of QRPA is considered by $\sum_{\pi \nu}$.
	In Fig.~\ref{Fig3}, we compare the sum of $|\delta_{N J^P}|^{2}$ for both mother and daughter nuclei of $\beta\beta$ decay.
	With the same kind of pairing, $\sum_{N J^P} |\delta_{N J^P}|^2$ for different Skyrme interactions are similar.
	However, for each Skyrme interaction, $\sum_{N J^P} |\delta_{N J^P}|^2$ obtained
	by the surface pairing is always larger than that obtained by the volume pairing.
	We notice that although the mean pairing gaps determined by the isovector pairing for the ground-state calculation are fixed to the experimental data,
	the occupations of single-particle levels
	around the Fermi surface are very different for volume-pairing and surface-pairing cases.
	In the case of the volume pairing, the distribution of occupation probability is much sharper than the case of surface pairing.
	As a result, there are more single-particle levels with partial occupations for the case of surface pairing,
	and this will lead to a larger 2qp space for QRPA calculation,
	and hence a larger $\sum_{N J^P} |\delta_{N J^P}|^2$ value, which immediately gives rise to a larger $M^{0\nu}$ when using surface pairing.
	
	The values of the overlaps $\langle {\rm HFB}_{f}|{\rm HFB}_{i} \rangle$ are similar for different pairing interactions for all considered nuclei,
	around 0.82, except for the semi-magic nucleus $^{136}$Xe, whose values are around $0.45$ and $0.25$ for the cases of volume pairing and surface pairing, respectively.
	This is because a sharper distribution of the occupation probability for neutrons in the daughter nucleus $^{136}$Ba
	in the case of volume pairing is more similar to
	that in the mother nucleus $^{136}$Xe, which is a step function due to the magic neutron number.
	As a result, although the $\sum_{N J^P} |\delta_{N J^P}|^2$ in $^{136}$Xe  is not different from other nuclei,
	by considering the overlap factor $\langle {\rm HFB}_{f}|{\rm HFB}_{i} \rangle$ for $^{136}$Xe,
	the $M^{0\nu}$ by different pairing interactions are similar,
	where volume pairing even gives a slightly larger $M^{0\nu}$ compared to the surface pairing case, as shown in Fig.~\ref{Fig1}.
	
	\begin{figure}[!t]
		\centering
		\includegraphics[width=0.48\textwidth]{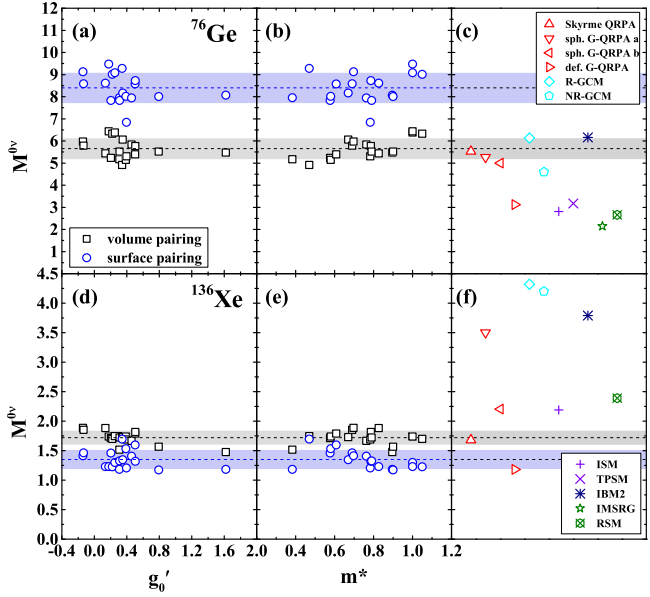}\\
		\caption{$M^{0\nu}$ for $^{76}$Ge [panels (a) and (b)] and $^{136}$Xe [panels (d) and (e)]
			with 18 Skyrme interactions and two kinds of pairing interactions.
			Blue circles and black squares are respectively the results of volume and surface pairing interactions.
			Their corresponding mean values and standard deviations are represented by the dashed lines and shaded regions.
			Results of other nuclear models for $^{76}$Ge and $^{136}$Xe are plot in panels (c) and (f), respectively.
		}\label{Fig4}
	\end{figure}
	
	Since $^{76}$Ge and $^{136}$Xe are of experimental interest
	to GERDA \cite{Agostini2020}, MAJORANA \cite{Aalseth2018}, and EXO \cite{Anton2019},
	KamLAND-Zen \cite{Gando2019}, XENON \cite{Aprile2022} collaborations,
	we further systematically examine the independence on $ph$ channel interaction
	and dependence on ground-state isovector $pp$ interaction of $M^{0\nu}$ for $^{76}$Ge and $^{136}$Xe.
	Besides the previously used SkM$^{\ast}$, SkO$'$, SLy4, and SGII,
	another 14 Skyrme interactions are employed.
	They are SLy5 \cite{Chabanat1998SLy45}, SKO \cite{Reinhard1999SkO}, SGI \cite{VanGiai1981SGII},
	SII, SIII, SIV \cite{Beiner1975SII},
	LNS \cite{Cao2006LNS}, SkT6 \cite{Tondeur1984SkT}, BSk1 \cite{Samyn2002BSk1}, MSk1 \cite{Tondeur2000MSk},
	SkI3, SkI4 \cite{Reinhard1995SkI3}, SAMi \cite{RocaMaza2012SAMi},
	and Z$_{\sigma}$ \cite{Friedrich1986Zsig}.
	The effective mass $m^{\ast}$ and Landau parameter $g'_0$ of these interactions span a wide range.
	The former will affect the structure of single-particle levels \cite{Vautherin1972},
	while the later plays an important role in spin-isospin excitations \cite{Fracasso2007}.
	Results of $M^{0\nu}$ with 18 Skyrme interactions and two kinds of pairing forces are depicted in Fig.~\ref{Fig4}.
	Their mean values $\overline{M}^{0\nu}$ and standard deviations $\sigma$ are listed in Tab.~\ref{Tab1}.
	For each kind of $pp$ interaction, $M^{0\nu}$ obtained by different $ph$ interactions are close, where
	the standard deviations $\sigma$ are only around 10\% of the mean values $\overline{M}^{0\nu}$.
	Besides, as the results in Fig.~\ref{Fig1}, $M^{0\nu}$ obtained by volume pairing are generally smaller
	for the open shell nucleus $^{76}$Ge, and larger for semi-magic nucleus $^{136}$Xe due to the occupation probability distribution around Fermi surface.
	
	We further make a brief comparison of the NMEs between our values and other theoretical results in Fig.~\ref{Fig4}(c) and (f).
	For $^{76}$Ge, our value of the NME is 5.65(45) in the case of volume pairing, which is very close to the previous Skyrme QRPA calculation \cite{Mustonen2013}.
	Also, the NMEs obtained by spherical $G$-QRPA \cite{Hyvaerinen2015,Simkovic2018},
	relativistic (R-) and non-relativistic (NR-) generator coordinate methods (GCM) \cite{Rodriguez2010,Yao2015},
	and interacting boson model (IBM2) \cite{Barea2013}
	lie within around $1.0\sim 2.0\sigma$ from our $\overline{M}^{0\nu}$ by volume pairing,
	while our results are about twice the NMEs obtained by the deformed $G$-QRPA \cite{Fang2018}, ISM \cite{Menendez2009},
	triaxial projected shell model (TPSM) \cite{Wang2021},
	and {\it ab initio} approaches including in-medium similarity renormalization group (IMSRG) \cite{Belley2021} and realistic shell model (RSM) \cite{Coraggio2020},
	which could be caused by the lack of complicated many-body correlations \cite{Simkovic2008,Engel2017,Yao2022a}.
	For $^{136}$Xe, our values of the NMEs are 1.72(11) by the volume pairing and 1.35(15) by the surface pairing,
	which are respectively close to the results of previous Skyrme QRPA and deformed $G$-QRPA.
	Either by the volume pairing or surface pairing, our results are smaller
	than the other models. 
	The reason could be the sharp neutron Fermi surface in $^{136}$Xe \cite{Mustonen2013}, which significantly suppresses the NMEs
	through the overlap of HFB functions.
	
	It is also noticed that in previous study with $G$-QRPA \cite{Rodin2003},
	the $M^{0\nu}$ values are independent of the form of different realistic nucleon-nucleon potentials
	after the particle-particle interaction is adjusted to correctly reproduce  the $2\nu\beta\beta$-decay rate,
	where the Coulomb corrected Woods-Saxon potential is used to calculate the ground-state single-particle energies
	while the residual interaction in QRPA is described by $G$-matrix \cite{FangThesis,Simkovic2004}.
	In this work, the same Skyrme interaction is consistently used for both ground-state and excited-state calculation
	in self-consistent QRPA. With completely different single-particle structure and residual $ph$ interaction
	obtained from a wide range of effective mass and Landau parameter, the $M^{0\nu}$ is still independent of the $ph$ interaction.
	In contrast, the form of $pp$ interaction is the main cause of uncertainties for $M^{0\nu}$ in self-consistent QRPA calculation.
	
	\begin{table}[b]
		\centering
		\caption{Mean values $\overline{M}^{0\nu}$ and the standard deviation $\sigma$ of $M^{0\nu}$
			obtained by 18 Skyrme interactions and two kinds of pairing interactions
			for $^{76}$Ge and $^{136}$Xe.}\label{Tab1}
		\renewcommand\arraystretch{1.3}
		\setlength{\tabcolsep}{4.55mm}{
			\begin{tabular}{ccccc}
				\hline \hline
				~                     & \multicolumn{2}{c}{Volume pairing} & \multicolumn{2}{c}{Surface pairing}     \\ \hline
				Nucleus                  & $^{76}$Ge & $^{136}$Xe & $^{76}$Ge & $^{136}$Xe \\ \hline
				$\overline{M}^{0\nu}$    &   5.65    &    1.72    &   8.40    &   1.35     \\ \hline
				$\sigma$ of ${M}^{0\nu}$ &   0.45    &    0.11    &   0.66    &   0.15     \\ \hline
				\hline
		\end{tabular}}
	\end{table}

	\begin{figure}[t]
		\centering
		\includegraphics[width=0.4\textwidth]{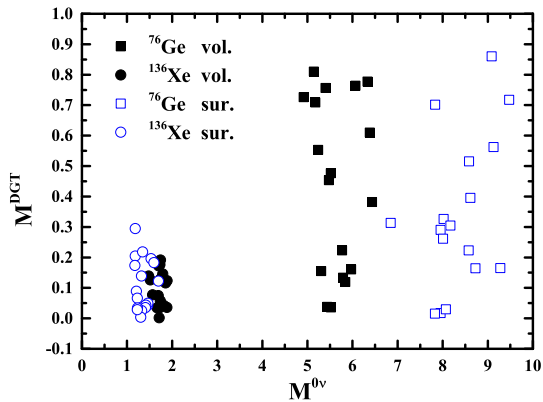}\\
		\caption{Double GT matrix elements $M^{\rm DGT}$ versus $M^{0\nu}$.
			Results are for two nuclei $^{76}$Ge (square) and $^{136}$Xe (circle) obtained by 18 Skyrme interactions and two kinds of pairing interactions, where solid and hollow symbols  are respectively the results of volume and surface pairing interactions.}\label{Fig5}
	\end{figure}
	
	In view of the recent interest in the study of the possible correlation
	between the matrix elements of double Gamow-Teller transition $M^{\rm DGT}$ and $M^{0\nu}$~\cite{Simkovic2018,Shimizu2018,Yao2022a},
	we examine this correlation with the QRPA using 18 Skyrme interaction in Fig.~\ref{Fig5}.
	Being contrary to the independence of $ph$ interaction in  $M^{0\nu}$,
	$M^{\rm DGT}$ is strongly affected by the choice of $ph$ interactions.
	Consequently, there seems no correlation between $M^{\rm DGT}$ and $M^{0\nu}$ in QRPA model.
	The reason could be the different behaviors of the distributions of $M^{\rm DGT}$ and $M^{0\nu}$ as functions of the inter-nucleon distance $r$ in QRPA.
	Both short range region ($r \lesssim 3$fm) and long range region ($r \gtrsim 3$fm) contribute much to $M^{\rm DGT}$,
	while only the short range region governs $M^{0\nu}$ \cite{Simkovic2011,Simkovic2018,Shimizu2018}.
	
	\section{Conclusions and Perspectives}
	\label{secConclu}
	In summary, the dependence of NMEs of $0\nu\beta\beta$ decay on $ph$ and $pp$ interactions is investigated
	in the framework of self-consistent QRPA based on Skyrme density functionals.
	Similar values for the $M^{0\nu}$ are obtained by different Skyrme interactions, namely, $ph$ interactions.
	In the systematical calculations of $\beta\beta$ emitters $^{76}$Ge and $^{136}$Xe with 18 Skyrme interactions having a large span over Landau parameter and effective mass,
	standard deviations are only around 10\% of the mean values $\overline{M}^{0\nu}$.
	However, $M^{0\nu}$ shows a dependence on $pp$ interaction.
	For open shell nuclei $^{76}$Ge, $^{82}$Se, $^{128}$Te, and $^{130}$Te,
	$M^{0\nu}$ obtained by surface pairing are always much larger than those obtained by volume pairing,
	due to the bigger 2qp space of QRPA model caused by a more smeared occupation probability distribution.
	The inverse case is found in the semi-magic nucleus $^{136}$Xe with much closer results, caused by the different situation in the calculation of
	overlap factor $\langle {\rm HFB}_{f}|{\rm HFB}_{i} \rangle$,
	where the sharp occupation probability distribution given by the volume pairing gives a larger value.
	We also investigate the correlation between $M^{\rm DGT}$ and $M^{0\nu}$.
	Due to the dependence of $M^{\rm DGT}$ on the $ph$ interaction, there seems no correlation between them.
	
	In the case of volume pairing, the NMEs for $^{76}$Ge and $^{136}$Xe are respectively
	5.65(45) and 1.72(11), while in the case of surface pairing, they are 8.40(66) and 1.35(15).
	The large uncertainty is originated from the isovector pairing interaction, which cannot be uniquely determined by the pairing gaps.
	Therefore, other constraints on the pairing interactions need to be considered
	in order to reduce the observed uncertainty in the NMEs by the QRPA method.
	
	\section*{Acknowledgement}
	Y.-F. Niu and W.-L. Lv acknowledge the support of the National Natural Science Foundation of China (Grant No. 12075104),
the “Young Scientist Scheme” of the National Key R\&D Program of China (Contract No. 2021YFA1601500),
and the Fundamental Research Funds for the Central Universities (Grant No. lzujbky-2021-it10).
D.-L. Fang acknowledges the support of the National Key R\&D Program of China (Contract No. 2021YFA1601300),
and the ``Light of West'' program and ``from zero to one" program by CAS.
J.-M. Yao is partially supported by the National Natural Science Foundation of China (Grant No. 12141501)
and the Fundamental Research Funds for the Central Universities, Sun Yat-sen University.
C.-L. Bai acknowledges the support of the National Natural Science Foundation of China (Grants No. 11575120 and No. 11822504).
J. Meng acknowledges the support of the National Key R\&D Program of China (Contract No. 2018YFA0404400), and
	the National Natural Science Foundation of China (Grants No. 11935003, No. 12070131001, and No. 12141501).

%

\end{document}